\documentclass[twocolumn,pra,aps,showpacs,amsmath,amssymb]{revtex4-1}
\usepackage{physics}
\usepackage{amsmath}
\usepackage{mathrsfs}
\usepackage{bbold}
\usepackage{graphicx}
\usepackage{color}
\usepackage{bm}
\usepackage{soul}
\usepackage{ulem}
\usepackage{dsfont}
\usepackage{bbm}
\usepackage{bigints}

\graphicspath{{NewFigures/}}

\usepackage{bigints}

\begin{document}

\title{Space-time-symmetric non-relativistic quantum mechanics: Time and position of arrival and an extension of a Wheeler–DeWitt-type equation}
\author{Eduardo O. Dias}
\email{eduardo.dias@ufpe.br}
\affiliation{Departamento de F\'{\i}sica, Centro de Ci\^encias Exatas e da Natureza, Universidade Federal de Pernambuco, Recife, PE  50670-901, Brazil}
	
\date{\today}

\begin{abstract}

We generalize a space-time-symmetric (STS) extension of non-relativistic quantum mechanics (QM) to describe a particle moving in three spatial dimensions. In addition to the conventional time-conditional (Schrödinger) wave function $\psi(x, y, z | t)$, we introduce space-conditional wave functions such as $\phi(t, y, z | x)$, where $x$ plays the role of the evolution parameter. The function $\phi(t, y, z | x)$ represents the probability amplitude for the particle to arrive on the plane $x = \text{constant}$ at time $t$ and transverse position $(y, z)$. Within this framework, the coordinate $x^\mu \in \{t, x, y, z\}$ can be conveniently chosen as the evolution parameter, depending on the experimental context under consideration. This leads to a unified formalism governed by a generalized Schrödinger-type equation,
$\hat{P}^{\mu} \ket{\phi^\mu(x^\mu)} = -i\hbar \, \eta^{\mu\nu} \frac{d}{dx^\nu} \ket{\phi^\mu(x^\mu)}$. It reproduces standard QM when $x^\mu = t$, with $|\phi^0(x^0)\rangle = |\psi(t)\rangle$, and recovers the STS extension when $x^\mu = x^i \in \{x, y, z\}$.  For a free particle, we show that $\phi(t, y, z | x) = \langle t, y, z | \phi(x) \rangle$ naturally reproduces the same dependence on the momentum wave function as the axiomatic Kijowski distribution. Possible experimental tests of these predictions are discussed. Finally, we demonstrate that the different states $|\phi^\mu(x^\mu)\rangle$ can emerge by conditioning (i.e., projecting) a timeless and spaceless physical state onto the eigenstate $|x^\mu\rangle$, leading to constraint equations of the form $\hat{\mathbbm{P}}^\mu |\Phi^\mu\rangle = 0$. This formulation generalizes the spirit of the Wheeler–DeWitt-type equation: instead of privileging time as the sole evolution parameter, it treats all coordinates on equal footing.

\end{abstract}

\pacs{Valid PACS appear here}
\maketitle

\section{\label{sec:level1}Introduction}

Asymmetries between space and time have long presented foundational challenges in quantum mechanics (QM), influencing not only the interpretation of the time-energy uncertainty relation~\cite{Time}, but also the prediction of time-of-arrival (TOA) distributions~\cite{Time,Time2,Muga,All,MugaComplex,Mielnik,Kijo2,Vona1,Leavens,Leavens4,Das2,Das,Das4,Vona} and attempts to unify QM with general relativity~\cite{Zeh,Anderson}. In standard QM, time is treated as a parameter, while position is a self-adjoint operator. Consequently, Born's rule gives a time-conditional (TC) character to the quantum state $|\psi(t)\rangle$, where $\psi(x|t)\equiv \langle x| \psi(t)\rangle$ is the amplitude of finding the particle at position $x$, given that a measurement occurs at time $t$.

This asymmetry complicates the prediction of time-related measurements within the orthodox postulates of QM. A key example is the TOA problem, which seeks the arrival-time distribution at a fixed position $x$, given an initial wave packet $\psi(x|t_0)$. Attempts to define a TOA operator ${\hat T}$~\cite{Time,Aharonov,Paul,Grot,Kijo,Galapon,Galapon1,Galapon2,Delgado} face the Pauli objection~\cite{Pauli}: requiring ${\hat T}$ to be self-adjoint and satisfy $[{\hat H},{\hat T}]=i\hbar$ implies that ${\hat H}$ is unbounded from below. Consequently, ideal models—those independent of the measurement process—must abandon either the commutation relation or the self-adjointness of ${\hat T}$~\cite{Delgado}.

Moreover, current TOA experiments are typically in the far-field or scattering regime, where semiclassical approximations provide accurate descriptions~\cite{Vona1,Feynman}. In this regime, different TOA models become experimentally indistinguishable, making it difficult to determine which distribution best captures the underlying physics. Combined with the Pauli objection, this has led to multiple inequivalent TOA models~\cite{Time,Muga}.

In this context, the Aharonov-Bohm TOA operator~\cite{Aharonov} has been supported from various perspectives~\cite{All,Grot,Kijo,Heger,Anastopoulos,HegerON1,HegerON2,Galapon}. Kijowski, for example, derived this operator by introducing a set of axioms that, in his view, a valid TOA distribution should satisfy within the framework of standard QM~\cite{Kijo}. However, these approaches face technical and interpretational issues~\cite{Mielnik,Kijo2,Vona1,Leavens,Leavens4,Das2,Vona}, such as the vanishing mean TOA for any real-valued $\psi(x|t_0)$ and violations of Kijowski's axioms by certain states. Another candidate for an ideal TOA distribution is the quantum flux density (i.e., the probability current)~\cite{MugaComplex,Leavens4,Grubl,Das,Das2,Das4,Vona1,Das3,Vona,Anantha,Leavens}. However, the backflow effect~\cite{Bracken,Bracken2,Halli}, where flux becomes negative, precludes its interpretation as a probability density and disqualifies it from being described by a POVM (positive-operator valued measure)~\cite{Vona}.

The conceptual difficulties associated with TOA extend to predicting the position at which a particle arrives on a detection screen~\cite{Kijo,Werner,Mielnik2}. For a detection plane located at $x = L$, the transverse arrival coordinates $(y, z)$ are registered at a random, uncontrollable time. Since $(y, z)$ generally depend on the TOA, they cannot be predicted using the fixed-time Schrödinger distribution $|\psi(L, y, z | t)|^2$~\cite{DasDSE}. Several models have addressed the space-time distribution of detection events~\cite{Schuss,Kudo,Wiseman,Kazemi,Kazemi2,Kazemi3}, including cumulative arrival distributions for the iconic double-slit experiment~\cite{DasDSE,Kudo,Kazemi2}. 

In this scenario, theoretical difficulties or internal inconsistencies in the existing approaches—combined with the absence of experimental confirmation—leave the TOA problem as an open question in quantum physics~\cite{Mielnik}. Since these issues arise within conventional QM, space-time-symmetric (STS) extensions such as Refs.~\cite{Piron,Dias,Dias3} offer an alternative framework to address the TOA problem. In particular, Refs.~\cite{Dias,Dias3} consider a one-dimensional spinless particle, treating time as a self-adjoint operator $\hat{t}$ in a complementary Hilbert space, with the Hamiltonian $\hat{h}$ defined via its canonical commutation relation with $\hat{t}$. Recent investigations~\cite{Dias2,Dias3,Lara,Lara2} have demonstrated promising applications of this STS framework, yielding consistent results for arrival and tunneling times~\cite{Dias2,Dias3,Lara} and an insightful derivation of its classical limit~\cite{Lara2}.

In this work, after briefly reviewing the one-dimensional STS extension proposed in Refs.~\cite{Dias,Dias3} (Sec.~\ref{review}), we extend this framework to three-dimensional motion in Sec.~\ref{STS3D}. There, we address both arrival time and arrival position on a surface using space-conditional (SC) wave functions, such as $\phi(t, y, z | x)$. In Sec.~\ref{Free}, we solve the corresponding free-particle equation and compare the resulting SC wave function with both the well-known Kijowski distribution~\cite{Kijo} and the conventional wave function of standard quantum mechanics. This comparison reveals key differences that may be probed experimentally, offering potential tests of the predictions of the SC framework. Finally, in Sec.~\ref{Unify}, we present a unified formulation that not only reconciles conventional QM with its three-dimensional STS extension but also proposes a generalization of the Wheeler--DeWitt-type equation, yielding a set of constraints that treats all coordinates on equal footing.

While preparing this work, the author recently came across an interesting note by Constantin Piron in Ref.~\cite{Piron}, presented by Louis de Broglie to the C. R. Acad. Sc. Paris, which proposes a formalism similar to the one developed here. In that brief note, Piron also introduces an SC wave function additional to the conventional Schrödinger wave function. However, as will be discussed at the end of Sec.~\ref{review}, the construction of his formalism differs from the approach introduced in Refs.~\cite{Dias,Dias3} and expanded upon in this manuscript. Moreover, although Piron's work presents appealing ideas, it is a concise note of just two and a half pages, offering only limited discussion and exploration of his formalism. In contrast, the present manuscript offers a detailed interpretation and analysis of the formalism, proposes alternative dynamical equations, explicitly solves the free-particle case with a discussion of possible experimental tests, and unifies the STS extension with conventional QM via a Wheeler–DeWitt-type equation—developments that go beyond those presented in Piron's work.

\section{Review of the STS extension of QM: one-dimensional formulation}
\label{review}

In this section, we provide a concise review of the STS extension of quantum mechanics, whose full formulation can be found in Refs.~\cite{Dias,Dias3}. It is worth emphasizing that the STS extension does not aim to replace conventional QM but to provide complementary predictions for certain experimental configurations.

In the STS extension for a one-dimensional, spinless particle, position $x$ becomes a parameter, while time is promoted to an operator ${\hat t}$ acting on a distinct Hilbert space ${\cal H}_T$. The operator ${\hat t}$ is canonically conjugate to the Hamiltonian ${\hat h}$,
\begin{equation}\label{time}
{\hat t}|t\rangle =t|t\rangle~~{\rm and}~~ [{\hat h}, {\hat t}]=i\hbar,
\end{equation}
where $|t\rangle \in {\cal H}_T$ and $\langle t|t' \rangle=\delta (t-t')$. Note that while ${\hat h}$ is defined by its commutation relation with ${\hat t}$, the Hamiltonian ${\hat H}$ in conventional QM is obtained by replacing $x$ and $p$ with the operators ${\hat x}$ and ${\hat p}$ (acting on ${\cal H}_X$) in a symmetrized version of the classical Hamiltonian. Consequently, the time-energy uncertainty relation $\Delta {\hat t} \Delta {\hat h} \geq \hbar/2$ naturally arises in the STS framework, just as the position-momentum uncertainty relation $\Delta{\hat x} \Delta {\hat p} \geq \hbar/2$ does in standard QM.

In the time representation, ${\hat h}$ takes the form
\begin{equation}\label{energy}
\langle t|{\hat h}|t'\rangle =i\hbar \frac{\partial}{\partial t}\delta(t-t'),
\end{equation}
and its eigenstate $|\varepsilon\rangle$ (${\hat h}|\varepsilon \rangle= \varepsilon |\varepsilon \rangle$) is given by
\begin{equation}\label{autoestadoH}
|\varepsilon\rangle =\frac{1}{\sqrt{2\pi \hbar}} \int_{-\infty}^{\infty}dt~ e^{-i\varepsilon t/\hbar} |t\rangle.
\end{equation}
Note that $|\varepsilon\rangle$ belongs to ${\cal H}_T$, whereas the energy eigenstate $|E\rangle$ in conventional QM (${\hat H}|E\rangle = E|E \rangle$) belongs to ${\cal H}_X$. The resolution of the identity in ${\cal H}_T$ reads
\begin{equation}\label{identityT}
\int_{-\infty}^{\infty} dt ~|t\rangle \langle t|=\int_{-\infty}^{\infty} d\varepsilon~ |\varepsilon \rangle\langle \varepsilon|= \mathbbm{1}.
\end{equation}

The physical state in the STS extension is conditioned on a given position $x$ and can be represented in the time or energy basis using Eq.~(\ref{identityT}),
\begin{equation}\label{expansionT1}
\ket{{\phi}(x)}=\int_{-\infty}^{\infty} dt~\phi(t|x) ~|t\rangle=\int_{-\infty}^{\infty} d\varepsilon ~{\bar \phi}(\varepsilon|x) ~|\varepsilon\rangle.
\end{equation}
Here, $\phi(t|x)=\langle t |\phi(x)\rangle$ [${\bar \phi}(\varepsilon|x)=\langle \varepsilon|\phi(x)\rangle$] is the probability amplitude for the particle to arrive at $x$ at time $t$ [or with energy $\varepsilon$]. Operationally, these are the probability amplitudes for $t$ and $\varepsilon$, given that the detector is located at $x$. Within this framework, the arrival time at a given position is treated with the same level of fundamental importance as the particle's position at a fixed time in conventional QM.

In this framework, the STS extension imposes that $|{\phi}(x)\rangle$ evolves in the spatial coordinate $x$, analogous to how $|\psi(t)\rangle$ evolves in time in conventional QM, thereby establishing a spatially conditional dynamics. Since momentum generates spatial translations, we apply the quantization rule of Eq.~(\ref{time}) to the classical expression $P(t,H;x)=\pm \sqrt{2m\big[H-V(x,t)\big]}$, yielding
\begin{eqnarray}\label{ruleP}
{\hat P}({\hat t},{\hat h};x) &=& \sigma_{z} \sqrt{2m\left[{\hat h} - V(x,{\hat t})\right]},
\end{eqnarray}
where $\sigma_z = \mathrm{diag}(+1,-1)$. Applying ${\hat P}$ to $|{\phi}(x)\rangle$ gives the one-dimensional “dynamic” equation of the STS extension,
\begin{equation}\label{SchroT}
{\hat P} \ket{{\phi}(x)} = -i\hbar \frac{d}{dx} \ket{{\phi}(x)}.
\end{equation}
Hereafter, we refer to this equation as the space-conditional (SC) Schrödinger equation. Note that lowercase letters are used to denote the fundamental operators, from which capital-letter operators, such as ${\hat P}$, are defined.

Projecting Eq.~(\ref{SchroT}) onto $|t\rangle$ yields
\begin{equation}\label{Schro2T}
\sigma_{z} \sqrt{2m\left(i\hbar \frac{\partial}{\partial t} - V(x,t)\right)} {\phi}(t|x) = -i\hbar \frac{\partial {\phi}(t|x)}{\partial x},
\end{equation}
which defines ${\phi}(t|x)$ as a two-component object,
\begin{equation}\label{solT}
{\phi}(t|x) = \mqty(\phi^{+}(t|x) \\ \phi^{-}(t|x)).
\end{equation}
Here, ${\phi}^+(t|x)$ (positive momentum) and ${\phi}^-(t|x)$ (negative momentum) are interpreted as representing arrivals from the left and right, respectively. By interchanging $x \rightleftarrows t$, the interpretation of Eq.~(\ref{Schro2T}) mirrors that of the conventional Schrödinger equation~\cite{Dias3}: Given an “initial” condition ${\phi}(t|x_0)$—which specifies the possible arrival times at position $x_0$—the solution ${\phi}(t|x)$ yields the probability amplitude for the particle to arrive at position $x$ at time $t$.

The probability that the particle arrives at $x$ within the interval $[t, t + dt]$ is given by
\begin{equation}\label{rhoT}
{\cal P}(t|x) dt = \frac{|\langle t|{\phi}(x)\rangle|^2}{\langle {\phi}(x)|{\phi}(x) \rangle} dt = \frac{{\phi}^\dagger(t|x) {\phi}(t|x)}{\int_{-\infty}^{\infty} dt |\phi(t|x)|^2} dt,
\end{equation}
where the dagger symbol denotes the Hermitian conjugate. The normalization $\langle {\phi}(x)|{\phi}(x) \rangle$ represents the total probability that the particle arrives at $x$, regardless of the arrival time. This normalization is necessary because ${\hat P}$ is not generally Hermitian—a reasonable feature: whereas a particle can be measured at any time (as captured by $\langle \psi(t)|\psi(t)\rangle=1$), its arrival at any specific position $x$ is not guaranteed, even if one waits indefinitely.

In the one-dimensional version of the approach in Ref.~\cite{Piron}, the proposed equation corresponding to Eq.~(\ref{Schro2T}) employs a sign $\pm$ instead of the Pauli matrix $\sigma_z$, so that the wave function is not a two-component object as in Eq.~(\ref{solT}). Moreover, Piron does not adopt the bra-ket notation used throughout this work. In his framework, a state $\Phi(t,x)$ is introduced as a spacetime-normalized wave function, subject to distinct dynamical equations depending on which variable is controlled in the experimental procedure. If the measurement is instantaneous, the time variable is precisely determined, and the particle’s state is treated as an eigenstate of the time operator, evolving according to the conventional Schrödinger equation. In this case, $\Phi(t,x)$ coincides with $\psi(x|t)$. Crucially, such time control constitutes a physical intervention that affects the system’s dynamics.

Conversely, when the measurement involves determining the arrival time at a known position, time becomes the uncertain quantity while space is controlled. This leads to a description in which $\Phi(t,x) = \phi(t|x)$ is treated as an eigenstate of the position operator and evolves according to a SC Schr\"odinger equation. According to Piron, these two descriptions are fundamentally incompatible, since the nature of the control variable—determined by the experimental context—directly influences the dynamics of the quantum state.

In contrast, the STS extension developed in this work offers a different perspective. The wave functions $\psi(t|x)$ and $\phi(t|x)$ represent intrinsic quantum states of the same particle, defined independently of any experimental setup. Nevertheless, they encode different statistical outcomes associated with specific measurement scenarios. Thus, the choice between using $\psi(x|t)$ or $\phi(t|x)$ depends on the experimental context—such as detecting arrival positions at a fixed time or arrival times at a fixed position. We argue that, when a measurement is described within conventional QM—taking into account the interaction with measuring devices such as detectors and clocks—the information encoded in $\phi(t|x)$ should be registered in the conventional quantum states of these devices in an appropriate ideal measurement limit. In this sense, $\phi(t|x)$ encodes the ideal arrival time of the particle at position $x$, just as $\psi(x|t)$ describes ideal position measurements at a given time.

In addition, unlike Ref.~\cite{Piron}, the approach developed in this manuscript does not regard $\psi(t|x)$ and $\phi(t|x)$ as particular eigenstates (of time and position, respectively) derived from a single spacetime-normalized wave function $\Phi(t,x)$. Instead, as discussed in Sec.~\ref{Unify}, these conditional states emerge from projecting a timeless and spaceless total state onto time or space eigenstates. These distinctions, along with other differences highlighted in the introduction, will be further examined in the sections that follow.

\section{The STS extension of QM: three-dimensional formulation}\label{STS3D}

\subsection{Theoretical framework and physical interpretation}

As discussed in the introduction, the generalization of the one-dimensional TOA problem to the three-dimensional case involves describing the arrival of a particle on a given surface (the detector screen). In Cartesian coordinates, the plane $x = \text{constant}$, for example, defines the detector screen, and $(t, y, z)$ represents the random space-time point of arrival. To describe this setup within the STS framework, we introduce the 3D wave function $\phi(t, y, z | x)$. In this formulation, the choice of state parameter depends on the arrival surface of interest. Different parameters yield distinct wave functions for the same particle. In this section, we focus on arrival at the plane $x = \text{constant}$; the adaptation to other Cartesian planes is straightforward.

In the 3D STS extension, $|\phi(x)\rangle$ still represents the particle’s state at position $x$, but the corresponding Hilbert space is enlarged to ${\cal H}_T \otimes {\cal H}_Y \otimes {\cal H}_Z$. The coordinate basis is defined as $\{|t, y, z\rangle\}$, and the quantization rules become
\begin{eqnarray}\label{TYZ}
{\hat t}|t,y,z\rangle = t|t,y,z\rangle~~&{\rm and}&~~ [{\hat t}, {\hat  h}]=-i\hbar,\nonumber\\
{\hat y}|t,y,z\rangle =y|t,y,z\rangle~~&{\rm and}&~~ [{\hat  y}, {\hat  p_y}]=i\hbar,\nonumber\\
{\hat  z}|t,y,z\rangle =z|t,y,z\rangle~~&{\rm and}&~~ [{\hat  z}, {\hat p}_z]=i\hbar,
\end{eqnarray}
where ${\hat  t}$, ${\hat  y}$, and ${\hat  z}$ represent ${\hat  t}\otimes \mathds{1}_Y \otimes \mathds{1}_Z $, $\mathds{1}_T\otimes  {\hat y} \otimes \mathds{1}_Z $, and $\mathds{1}_T\otimes \mathds{1}_Y \otimes {\hat z} $, respectively. Additionally, $\langle t,y,z|t',y',z' \rangle = \delta (t-t')\delta (y-y')\delta (z-z')$. Although this quantization rule for $y$ and $z$ is equivalent to that of conventional QM, we will verify that the information about $y$ and $z$ encoded in $|\phi(x)\rangle$ is distinct from, and complementary to, that encoded in $|\psi(t)\rangle$.

Equation~(\ref{TYZ}) leads to the uncertainty relations
\begin{eqnarray}\label{uncertainty}
\Delta {\hat  t} \Delta {\hat h} \geq \hbar/2,~\Delta {\hat  y} \Delta {\hat  p_y} \geq \hbar/2,~{\rm and}~\Delta {\hat  z} \Delta {\hat  p}_z \geq \hbar/2,
\end{eqnarray}
and yields the coordinate representations given by
\begin{eqnarray}\label{EPyPz}
\langle t,y,z|{\hat  h}|t',y',z'\rangle  &=&i\hbar \frac{\partial}{\partial t}\delta(t-t')\delta(y-y')\delta(z-z'),\nonumber\\
\langle t,y,z|{\hat p}_y|t',y',z'\rangle  &=&-i\hbar \delta(t-t') \frac{\partial}{\partial y}\delta(y-y')\delta(z-z'),\nonumber\\
\langle t,y,z|{\hat  p}_z|t',y',z'\rangle  &=&-i\hbar \delta(t-t')\delta(y-y') \frac{\partial}{\partial z}\delta(z-z').\nonumber\\
\end{eqnarray}
The eigenstates of the conjugate momenta (${\hat h}|\varepsilon\rangle= \varepsilon |\varepsilon\rangle$, ${\hat  p}_y|p_y\rangle= p_y |p_y\rangle$, and ${\hat p}_z|p_z\rangle= p_z |p_z\rangle$) are represented in the coordinate basis as 
\begin{eqnarray}\label{autoestadoHPyPz}
|\varepsilon\rangle&=&\frac{1}{\sqrt{2\pi \hbar}} \int_{-\infty}^{\infty}dt~ e^{-iE t/\hbar} |t\rangle,\nonumber\\
|p_y\rangle&=&\frac{1}{\sqrt{2\pi \hbar}} \int_{-\infty}^{\infty}dy~ e^{ip_y y/\hbar} |y\rangle,\nonumber\\
|p_z\rangle&=&\frac{1}{\sqrt{2\pi \hbar}} \int_{-\infty}^{\infty}dz~ e^{ip_z z/\hbar} |z\rangle
\end{eqnarray}
Similarly to the one-dimensional case, in Eq.~(\ref{autoestadoHPyPz}), $|\varepsilon\rangle \in {\cal H}_T $, whereas in conventional QM, the energy eigenstate $|E\rangle \in {\cal H}_X \otimes {\cal H}_Y \otimes  {\cal H}_Z$.

When expressing $|\phi(x)\rangle$ in the position or momentum representation, we obtain  
\begin{eqnarray}\label{expansionTYZ}  
|\phi(x)\rangle &=& \int dtdydz ~\phi(t,y,z|x) |t,y,z\rangle \nonumber\\
&=& \int d\varepsilon dp_ydp_z ~\tilde{\phi}(\varepsilon,p_y,p_z|x) |\varepsilon,p_y,p_z\rangle,\nonumber\\
\end{eqnarray}  
respectively, with $\phi(t,y,z|x)$ and $\tilde{\phi}(\varepsilon,p_y,p_z|x)$ being two-component objects. Here $\phi(t,y,z|x)$ [$\bar{\phi}(\varepsilon,p_y,p_z|x)$] is interpreted as the probability amplitude for the particle to arrive on the plane $x = \text{constant}$ at the space-time point $(t,y,z)$ [or with energy and transverse momenta $(\varepsilon,p_y,p_z)$]. The total probability for the particle to arrive at $x = \text{constant}$, regardless of $(t,y,z)$ [or $(\varepsilon,p_y,p_z)$], is given by
\begin{eqnarray}\label{normEPyPz}
\langle \phi(x)|\phi(x)\rangle = \sum_{r=+,-} \int dtdydz ~|\phi^r(t,y,z|x)|^2\nonumber\\ 
=\sum_{r=+,-} \int dp_ydp_zd\varepsilon ~ |{\bar {\phi}}^r(\varepsilon,p_y,p_z|x)|^2.
\end{eqnarray}
Thus, Eq.~(\ref{rhoT}) generalizes to  
\begin{eqnarray}\label{rhoTYZ}  
{\cal P}_\phi (t,y,z|x) &=& \frac{|\langle t,y,z|{ \phi}(x)\rangle|^2}{\langle {\phi}(x)|{ \phi}(x) \rangle} \nonumber\\  
&=& \frac{{\phi}^{\dagger}(t,y,z|x){\phi}(t,y,z|x)}{\int_{-\infty}^{\infty} dtdydz ~|\phi(t,y,z|x)|^2}, 
\end{eqnarray}  
representing the probability for the particle to arrive on $x = \text{constant}$ within the space-time volume $dtdydz$. In this context, a detector must cover the entire $x = \text{constant}$ plane and remain continuously active, awaiting the particle’s arrival.

Equation~(\ref{SchroT}) continues to govern the dependence of $|\phi(x)\rangle$ on $x$, but now ${\hat P}$ is defined by applying the quantization rule of Eq.~(\ref{TYZ}) to the three-dimensional momentum in the $x$-direction:
\begin{eqnarray}\label{rulePx}
&&P_x=\pm \sqrt{2m\big [H-V(t,x,y,z)\big ]-p_y^2 -p_z^2}  \nonumber\\
&\rightarrow &  {\hat P}_x =\sigma_{z} \sqrt{2m\left[{\hat  h}-{\hat  V}({\hat  t},x,{\hat  y},{\hat  z})\right]- {\hat  p}_y^2- {\hat  p}_z^2}.\nonumber\\
\end{eqnarray}
When projected onto the basis $\ket{t,y,z}$, Eq.~(\ref{SchroT}) becomes
\begin{eqnarray}\label{Schro2TYZ}
\sigma_{z}\sqrt{2m\left[i\hbar \pdv{t}-V(t,{\vec x})\right]+\hbar^2 \left[\frac{\partial^2}{\partial y^2}+ \frac{\partial^2}{\partial z^2}\right]} { \phi}(t,y,z|x)\nonumber\\
=-i\hbar \frac{\partial}{\partial x}{ \phi}(t,y,z|x),\nonumber\\
\end{eqnarray}
where ${\vec x} = (x,y,z)$. The SC wave function remains a two-component object, given by
\begin{equation}\label{solT3D}
{\phi}(t,y,z|x)=\mqty(\phi^{+}(t,y,z|x)  \\  \phi^{-}(t,y,z|x)).
\end{equation}
The interpretation of Eq.~(\ref{Schro2TYZ}) is analogous to that of Eq.~(\ref{Schro2T}): given an “initial” amplitude $\phi(t,y,z|x_0)$, which encodes the probability of arrival on the plane $x = x_0$, Eq.~(\ref{Schro2TYZ}) determines the amplitude $\phi(t,y,z|x)$ corresponding to arrival on an arbitrary plane $x = \text{constant}$.

\subsection{Alternative formulation of the momentum operator}

Simply requiring that ${\hat P}_x$ satisfies the classical dispersion relation does not uniquely determine its form. In contrast to the prescription used in Eq.~(\ref{rulePx}), ${\hat P}_x$ can instead be defined by applying the Dirac procedure from conventional relativistic quantum mechanics. This approach was implemented in Ref.~\cite{Lara2} for the one-dimensional case. In this section, we adopt a version of ${\hat P}_x$ that, in the one-dimensional limit, differs from the operator proposed in Ref.~\cite{Lara2}. The physical implications of this freedom are discussed below.

For a time-independent potential, instead of using $\sigma_z$ to encode positive and negative momentum components, we can define ${\hat P}_x$ as
\begin{eqnarray}\label{Dirac}
{\hat P}_x=  \alpha \sqrt{2m{\hat  h}} +\beta \sqrt{-2mV(x,{\hat y},{\hat z})-{\hat  p}_y^2-{\hat p}_z^2},
\end{eqnarray}
where $\alpha$ and $\beta$ are matrices to be determined. In order for ${\hat P}_x$ to satisfy the non-relativistic dispersion relation, we require
\begin{eqnarray}\label{DiracCondition}
\alpha^2~ 2m{\hat h} + \beta^2~[-2mV(x,{\hat y},{\hat z})-{\hat  p}_y^2-{\hat p}_z^2] \nonumber\\
 ( \alpha \beta +\beta \alpha) ~ \sqrt{2m{\hat  h}} \sqrt{-2mV(x,{\hat y},{\hat z})-{\hat  p}_y^2-{\hat p}_z^2} \nonumber\\
=2m{\hat  h}- 2mV(x,{\hat y},{\hat z})-{\hat  p}_y^2-{\hat p}_z^2,
\end{eqnarray}
where we have used the fact that $\hat h$ commutes with $V(x,{\hat y},{\hat z})$, ${\hat p}_y$, and ${\hat p}_z$. From Eq.~(\ref{DiracCondition}), we find that $\alpha^2 = \mathbbm{1}$, $\beta^2 = \mathbbm{1}$, and $\alpha\beta + \beta\alpha = 0$. An operator ${\hat P}_x$ that fulfills these conditions is given by
\begin{eqnarray}\label{rulePx2}
{\hat P}_x=  \sigma_z \sqrt{2m{\hat  h}} +i\sigma_x \sqrt{2mV(x,{\hat y},{\hat z})+{\hat  p}_y^2+{\hat p}_z^2},
\end{eqnarray}
where $\sigma_x$ is the Pauli matrix in the $x$-direction. In Ref.~\cite{Lara2}, the authors studied the one-dimensional case, i.e., ${\hat p}_y^2 = {\hat p}_z^2 = 0$, and, unlike Eq.~(\ref{rulePx2}), chose $\alpha = \sigma_x$ and $\beta = \sigma_z$. Our choice is justified below.

A consequence of this freedom is that the resulting SC Schr\"odinger equation depends on the specific definition of ${\hat P}_x$.An appealing feature of Eq.~(\ref{rulePx2}) is that, unlike Eq.~(\ref{rulePx}) and the operator ${\hat P}_x$ defined in Ref.~\cite{Lara2}, it introduces coupling between $\phi^+$ and $\phi^-$ only in the presence of an interaction. In this formulation, as $\phi^+$ enters a region where $V \neq 0$, it transfers amplitude to $\phi^-$, and vice versa. This behavior is consistent with standard QM, where the Schrödinger wave function undergoes reflection at interfaces between regions with different potentials. Consistent with this reasoning, another possible choice for ${\hat P}_x$ [although restricted to the case $V=V(x)$] is ${\hat P}_x=  \sigma_z \sqrt{2m{\hat  h}-{\hat  p}_y^2-{\hat p}_z^2} +i\sigma_x \sqrt{2mV(x)}$.

Substituting Eq.~(\ref{rulePx2}) into Eq.~(\ref{SchroT}), we find that $\phi^+$ and $\phi^-$ satisfy the coupled equations
\begin{equation}\label{rulePx3}
\begin{cases}
       \sqrt{i\hbar\dfrac{\partial}{\partial t}}\phi^+ +i \sqrt{V-\hbar^2\left(\dfrac{d^2 }{dx^2}+\dfrac{d^2 }{dy^2}\right)}\phi^-=-\dfrac{i\hbar}{\sqrt{2m}} \dfrac{\partial \phi^+}{\partial x}\\
       \sqrt{i\hbar\dfrac{\partial}{\partial t}}\phi^- -i \sqrt{V-\hbar^2\left(\dfrac{d^2 }{dx^2}+\dfrac{d^2 }{dy^2}\right)}\phi^+=\dfrac{i\hbar}{\sqrt{2m}} \dfrac{\partial \phi^-}{\partial x}.
    \end{cases}
\end{equation}
To derive from Eq.~(\ref{rulePx3}) an expression analogous to the stationary Schr\"odinger equation in conventional QM, we begin by considering the three-dimensional version of Eq.~(\ref{expansionT1}):
\begin{equation} \label{expansionE3D}
    |\phi(x)\rangle=\int d\varepsilon ~dydz~{\bar \phi}(\varepsilon,y,z|x) ~|\varepsilon,y,z\rangle.
\end{equation}
Here, ${\bar \phi}(\varepsilon,y,z|x)$ is a two-component object representing the probability amplitude for the particle to arrive on the plane $x = \text{constant}$ at position $(y,z)$ with energy $\varepsilon$. Projecting Eq.~(\ref{expansionE3D}) onto $|t,y,z\rangle$, each component becomes
\begin{equation}  \label{expansionE3D2}
    \phi^\pm(t,y,z|x)=\frac{1}{\sqrt{2 \pi \hbar}} \int d\varepsilon ~{\bar \phi}^\pm(\varepsilon,y,z|x) ~e^{-i\varepsilon t/\hbar}.
\end{equation}

It is worth noting that the counterpart of Eq.~(\ref{expansionE3D2}) in conventional QM is the wave function expressed as a superposition of stationary states,
\begin{eqnarray}
    \psi(x,y,z|t)&=& \int dE ~{\bar \psi}_E(x,y,z|t) ~e^{-iE t/\hbar}.
\end{eqnarray}
Unlike the interpretation of ${\bar \phi}(\varepsilon,y,z|x)$, the quantity ${\bar \psi}_E(x,y,z|t)$ represents the probability amplitude that, at time $t$, the particle is detected at position $(x,y,z)$, given that its energy is $E$. Thus, whereas in ${\bar \psi}_E(x,y,z|t)$ the energy $E$ is a conditional variable, $\varepsilon$ is a probabilistic variable in Eq.~(\ref{expansionE3D2}).

The stationary SC Schr\"odinger equation is obtained by substituting Eq.~(\ref{expansionE3D2}) into Eq.~(\ref{rulePx3}) and identifying $\sqrt{d/dt}$ as the Riemann-Liouville fractional derivative $_{-\infty}D^{1/2}|_t$~\cite{Dias,Dias2,Dias3,Lara}, which is equivalent to the Caputo fractional derivative~\cite{frac}. Using the relation $\sqrt{i\hbar d/dt}~ \exp\{-i \varepsilon t/\hbar\} =\sqrt{i\hbar} _{-\infty}D^{1/2}_t \exp\{-i \varepsilon t/\hbar\} = \sqrt{\varepsilon} \exp\{-i \varepsilon t/\hbar\}$, Eq.~(\ref{rulePx3}) can be rewritten in terms of ${\bar \phi}^\pm = {\bar \phi}^\pm (\varepsilon, y, z | x)$ as
\begin{equation}\label{rulePx4}
\begin{cases}
       \sqrt{\varepsilon}{\bar \phi}^+ + i\sqrt{V-\hbar^2\left(\dfrac{d^2 }{dx^2}+\dfrac{d^2 }{dy^2}\right)}{\bar \phi}^- = -\dfrac{i\hbar}{\sqrt{2m}} \dfrac{\partial {\bar \phi}^+}{\partial x}\\
       \sqrt{\varepsilon}{\bar \phi}^- -i \sqrt{V-\hbar^2\left(\dfrac{d^2 }{dx^2}+\dfrac{d^2 }{dy^2}\right)}{\bar \phi}^+= \dfrac{i\hbar}{\sqrt{2m}} \dfrac{\partial {\bar \phi}^-}{\partial x}.
    \end{cases}
\end{equation}

For the one-dimensional case, where $V = V(x)$, the components ${\bar \phi}^+$ and ${\bar \phi}^-$ in Eq.~(\ref{rulePx4}) can be decoupled by isolating ${\bar \phi}^-$ (${\bar \phi}^+$) in the first (second) equation and substituting it into the second (first) one. The resulting expressions for ${\bar \phi}^\pm = {\bar \phi}^\pm(\varepsilon|x)$ are
\begin{eqnarray}\label{rulePx5}
&&-\frac{\hbar^2}{2m}\left(\frac{d^2 {\bar \phi}^\pm }{dx^2}-\frac{1}{2V}\frac{dV}{dx} \frac{d {\bar \phi}^\pm}{dx} \right)\nonumber\\
&&+\left(V \mp i\hbar \sqrt{\frac{\varepsilon}{2m}}\frac{1}{2V}\frac{dV}{dx}\right) {\bar \phi}^\pm =\varepsilon {\bar \phi}^\pm.
\end{eqnarray}
Comparing Eq.~(\ref{rulePx5}) with the stationary Schrödinger equation, we observe two additional contributions: one proportional to $d{\bar \phi}^\pm/dx$, and another that gives rise to an effective potential, represented by the term within the second parentheses. In Ref.~\cite{Lara2}, the resulting SC Schrödinger equation includes only a single correction term, which likewise yields an effective potential similar to that in Eq.~(\ref{rulePx5}). A more detailed investigation of the effects of the different definitions of ${\hat P}_x$ on the SC Schr\"odinger equation is left for future work.

\subsection{Comparative analysis between $\psi(x,y,z|t)$ and $\phi(t,y,z|x)$ }

The classical version of conventional QM and its STS extension can be represented, respectively, by
\begin{eqnarray} \label{Comparison}
{\rm QM}: (t,x(t),y(t),z(t)) \quad {\rm and} \quad {\rm STS}: (t(x),x,y(x),z(x)). \nonumber\\
\end{eqnarray}
While both representations in Eq.~(\ref{Comparison}) describe the same underlying physics, in this section we qualitatively examine how their quantum counterparts, $\psi(x,y,z|t)$ and $\phi(t,y,z|x)$, exhibit fundamentally distinct features. In particular, we focus on the predictions associated with the $(y, z)$ coordinates, which naturally arise in the three-dimensional generalization of Ref.~\cite{Dias,Dias3}.

For simplicity, we restrict the analysis to motion in the $(x,y)$ plane. We begin by considering the following cumulative distributions:
\begin{equation}\label{Ppsi}
{\cal P}_\psi(y|t):=\int dx~|\psi(x,y|t)|^2
\end{equation}
and
\begin{equation}\label{Pphi}
{\cal P}_\phi(y|x):=\int dt~|\phi(t,y|x)|^2,
\end{equation}
where we assume $\langle \phi(x)|\phi(x)\rangle=1$. While ${\cal P}_\psi(y|t)$ represents the probability of measuring $y$ at a fixed instant $t$, independently of the measured $x$ coordinate, ${\cal P}_\phi(y|x)$ gives the probability of measuring $y$ at a fixed position $x$, regardless of the time $t$ at which the measurement occurs. 

It is important to emphasize that, beyond representing distinct conditional probabilities, ${\cal P}_\psi(y|t)$ and ${\cal P}_\phi(y|x)$ also correspond to physically distinct experimental setups. In the experiment associated with Eq.~(\ref{Pphi}), the detectors are placed along the plane $x = L$ and remain continuously active over time. In contrast, to properly measure Eq.~(\ref{Ppsi}), the detectors must be triggered precisely at a predetermined instant $t$, with the measurement occurring within a very short time interval $\Delta t \rightarrow 0$. Furthermore, the detectors must cover the entire $(x,y)$ plane.

Another relevant comparison between the probability distributions assigned to $y$ arises when considering that, in both formalisms, the detectors are exclusively distributed along the plane $x = L$, with $L$ constant. In this case, we investigate the probability density
\begin{equation}\label{PpsiL}
{\cal P}_\psi(y|t;L) := \frac{|\psi(L,y|t)|^2}{\int dy\, |\psi(L,y|t)|^2}
\end{equation}
obtained from conventional QM, and the probability density within the STS extension given by
\begin{equation}\label{PphiL}
{\cal P}_\phi(y|t;L) := \frac{{\cal P}_\phi(t,y|L)}{{\cal P}_\phi(t|L)} = \frac{|\phi(t,y|L)|^2}{\int dy\, |\phi(t,y|L)|^2}.
\end{equation}

At first glance, one might expect that both Eqs.~(\ref{PpsiL}) and~(\ref{PphiL}) describe the probability of a particle arriving on the plane $x = L$ at the coordinate $y$ and at time $t$. However, this prediction is actually provided only by Eq.~(\ref{PphiL}), which corresponds to an experimental setup where the detector is permanently located at $x = L$, passively waiting for the particle’s arrival, as previously discussed. In this case, only the detection events that occurred precisely at time $t$ are post-selected. By contrast, Eq.~(\ref{PpsiL}) corresponds to a scenario in which the detector is positioned at $x = L$ only at the instant $t$, during a narrow time window $\Delta t \to 0$. Under such fundamentally different detection procedures, it is not evident that the two distributions should yield equivalent predictions, even in an approximate or limiting regime.

Therefore, from the perspective of the 3D STS extension of QM, the predictions for the coordinates $y$ and $z$ in both formalisms provide complementary information about the same particle. This reasoning extends to any observable that depends on combinations of the transverse coordinates $y$ and $z$.

\section{The free-particle case in the three-dimensional STS extension}\label{Free}

In this section, we explore the predictions of the STS extension for a free particle, adopting the original definition of ${\hat P}_x$ given in Eq.~(\ref{rulePx}). To solve Eq.~(\ref{Schro2TYZ}) for $V(t,{\vec x})=0$, we apply $-i\hbar \partial/\partial x$ to both sides of Eq.~(\ref{Schro2TYZ}), obtaining
\begin{eqnarray}\label{Schro1TYZfree}
\sqrt{2m i\hbar \pdv{t}+\hbar^2 \left[\frac{\partial^2}{\partial y^2}+ \frac{\partial^2}{\partial z^2}\right]} \left(-i\hbar \frac{\partial}{\partial x} \right){ \phi}^\pm(t,y,z|x)\nonumber\\
=-\hbar^2 \frac{\partial^2}{\partial x^2}{ \phi}^\pm(t,y,z|x).~~
\end{eqnarray}
Here we have used the fact that the derivative with respect to $x$ commutes with ${\hat P}_x$ in the free-particle case ($V=0$). Replacing $-i\hbar \partial \phi/\partial x$ on the left-hand side of Eq.~(\ref{Schro1TYZfree}) by the left-hand side of Eq.~(\ref{Schro2TYZ}) [with $V=0$], and dividing both sides by $2m$, we recover the Schrödinger equation:
\begin{eqnarray}\label{Schro2TYZfree}
 \left[i\hbar \pdv{t}+\frac{\hbar^2}{2m} \left(\frac{\partial^2}{\partial y^2}+ \frac{\partial^2}{\partial z^2}\right)\right]{ \phi}^\pm(t,y,z|x) \nonumber\\
=-\frac{\hbar^2}{2m} \frac{\partial^2}{\partial x^2}{ \phi}^\pm(t,y,z|x).
\end{eqnarray}

Although the STS extension leads to the Schr\"odinger equation in the particular case of a free particle, the ``initial'' condition of ${ \phi}^\pm(t,y,z|x)$ and its temporal normalization make it distinct from $\psi(x,y,z|t)$. It is worth noting that ${ \phi}^\pm(t,y,z|x)$ does not obey the Schrödinger equation when the potential depends on $x$, since $[-i\hbar \partial/\partial x, {\hat P}_x] \neq 0$.

With $\phi^+$ ($\phi^-$) describing a particle arriving at $x$ from the right (left) with positive (negative) momentum, a particular solution of Eq.~(\ref{Schro2TYZfree}) is
\begin{eqnarray}\label{SolTYZ1}
{\phi}^{\pm}(t,y,z|x)=\frac{1}{(2\pi\hbar)^{3/2}} ~e^{i(\pm |p_x| x+ p_y y+ p_z z - \varepsilon t)/\hbar },
\end{eqnarray}
with $\varepsilon=(p_x^2+p_y^2+p_z^2)/(2m)$. Using this dispersion relation to express $p_x$ in terms of $\varepsilon$, $p_y$, and $p_z$, Eq.~(\ref{SolTYZ1}) becomes
\begin{eqnarray}\label{SolTYZ2}
{\phi}^{\pm}(t,y,z|x)&=&\frac{1}{(2\pi\hbar)^{3/2}} 
 ~e^{i(p_y y+ p_z z - \varepsilon t)/\hbar } \nonumber\\
 &\times& e^{\pm i\sqrt{2m\varepsilon - p_y^2 - p_z^2}~x/\hbar}.
\end{eqnarray}
As a result, the general solution of Eq.~(\ref{Schro2TYZ}) for a free particle can be expressed as
\begin{eqnarray}\label{SolTYZ}
{\phi}^{\pm}(t,y,z|x)=\frac{1}{(2\pi\hbar)^{3/2}}\int_{-\infty}^{\infty}  dp_y dp_z \int_{(p_y^2+p_z^2)/2m}^{\infty} d\varepsilon~\nonumber\\
A^{\pm}(\varepsilon,p_y,p_z) ~e^{i(p_y y + p_z z - \varepsilon t)/\hbar}~e^{\pm i\sqrt{2m \varepsilon - p_y^2 - p_z^2}~x/\hbar},\nonumber\\
\end{eqnarray}
where $p_x$ is real, which implies that $\varepsilon \geq (p_y^2 + p_z^2)/(2m)$.

The physical interpretation of the coefficients $A^\pm (\varepsilon,p_y,p_z)$ emerges from evaluating $|\phi(x)\rangle$. Substituting Eq.~(\ref{SolTYZ}) into Eq.~(\ref{expansionTYZ}), and then integrating first over $t$, $y$, and $z$, we obtain
\begin{eqnarray}\label{SolTYZ4}
&&|\phi(x)\rangle = \int_{-\infty}^{\infty}  dp_ydp_z \int_{(p_y^2+p_z^2)/2m}^{\infty}d\varepsilon ~A^{\pm}(\varepsilon,p_y,p_z) \nonumber\\
&&\times~e^{\pm i\sqrt{2m\varepsilon- p_y^2 -p_z^2}x/\hbar}
 \int_{-\infty}^{\infty} dt dydz ~\frac{e^{i(p_y y+ p_z z- \varepsilon t)/\hbar}}{(2\pi\hbar)^{3/2}}|t,y,z\rangle.\nonumber\\
\end{eqnarray}
On the other hand, from Eq.~(\ref{autoestadoHPyPz}), we know that
\begin{equation}\label{projectionTYZ2}
|\varepsilon,p_y,p_z \rangle= \int_{-\infty}^{\infty} dt dydz ~\frac{e^{i(p_y y+ p_z z- \varepsilon t)/\hbar}} {(2\pi\hbar)^{3/2}}|t,y,z\rangle.
\end{equation}
Substituting Eq.~(\ref{projectionTYZ2}) into Eq.~(\ref{SolTYZ4}) results in
\begin{eqnarray}\label{SolTYZ5}
|\phi(x)\rangle &=& \int_{-\infty}^{\infty} d\varepsilon  dp_ydp_z  ~A^{\pm}(\varepsilon,p_y,p_z) \nonumber\\
&\times& e^{\pm i\sqrt{2m\varepsilon- p_y^2 -p_z^2}x/\hbar}~|\varepsilon,p_y,p_z \rangle.
\end{eqnarray}
Comparing Eqs.~(\ref{SolTYZ5}) and~(\ref{expansionTYZ}), we immediately identify
\begin{eqnarray}\label{coef1}
{\bar \phi}^\pm(\varepsilon,p_y,p_z|x)= ~A^{\pm}(\varepsilon,p_y,p_z) ~e^{\pm i\sqrt{2m\varepsilon- p_y^2 -p_z^2}x/\hbar}.
\end{eqnarray}
Thus,
\begin{eqnarray}\label{coef2}
A^{\pm}(\varepsilon,p_y,p_z) = {\bar \phi}^{\pm}(\varepsilon,p_y,p_z|x=0):={\bar \phi}^{\pm}(\varepsilon,p_y,p_z),\nonumber\\
\end{eqnarray}
representing the probability amplitude for a free particle to arrive on the plane $x=0$ with energy $\varepsilon$ and momenta $p_y$ and $p_z$.

We now aim to compare the general solution of Eq.~(\ref{SolTYZ}) with both the axiomatic Kijowski distribution and the free-particle wave function from standard QM. To this end, we express Eq.~(\ref{SolTYZ}) in terms of the momentum wave function ${\tilde \phi}(p_x, p_y, p_z | x)$, which represents the probability amplitude for a particle to arrive at position $x$ with momentum $\vec{p} = (p_x, p_y, p_z)$. This momentum wave function can be related to the coefficients ${\bar \phi}^\pm(\varepsilon, p_y, p_z | x)$ introduced in Eq.~(\ref{coef1}) by means of the normalization condition in Eq.~(\ref{normEPyPz}). 

Since $p_y$ and $p_z$ are fixed in the integration over energy $\varepsilon$ in Eq.~(\ref{normEPyPz}), we may change the integration variable from $\varepsilon$ to $p_x$, using the dispersion relation $\varepsilon = (p_x^2 + p_y^2 + p_z^2)/2m$ and the differential identity
\begin{equation}\label{change}
d\varepsilon\big|_{p_y, p_z} = \frac{\partial \varepsilon}{\partial p_x} \, dp_x = \frac{p_x}{m} \, dp_x.
\end{equation}
Substituting Eq.~(\ref{change}) into Eq.~(\ref{normEPyPz}), we obtain
\begin{eqnarray}\label{normPxPyPz2}
&&\langle \phi(x)|\phi(x)\rangle = \sum_{r=\pm} \int_{-\infty}^{\infty}dp_y~dp_z \int_{0}^{\infty}dp_x\nonumber\\
&&\times~\left |\sqrt{\frac{p_x}{m}} ~{\bar {\phi}}^r (\varepsilon=(p_x^2+p_y^2+p_z^2)/2m,p_y,p_z|x)\right|^2.\nonumber\\
\end{eqnarray}
Here, we use the fact that in Eq.~(\ref{SolTYZ}), the integration over $\varepsilon$ ranges from $(p_y^2 + p_z^2)/2m$ to $\infty$, which implies that the corresponding domain for $p_x$ is $[0, \infty)$. We therefore identify the probability density $|{\tilde{\phi}}(p_x,p_y,p_z|x)|^2$ with the integrand of Eq.~(\ref{normPxPyPz2}), leading to
\begin{eqnarray}\label{con}
&&{\tilde {\phi}}(\pm|p_x|,p_y,p_z|x)=\nonumber\\
&&\sqrt{\frac{|p_x|}{m}}~{\bar {\phi}}^\pm (\varepsilon=(p_x^2+p_y^2+p_z^2)/2m,p_y,p_z|x).
\end{eqnarray}
Using Eq.~(\ref{coef1}), Eq.~(\ref{con}) can be rewritten as
\begin{eqnarray}\label{Amp3D}
{\tilde {\phi}}(p_x,p_y,p_z|x)={\tilde {\phi}}(p_x,p_y,p_z) ~e^{ip_x x/\hbar},
\end{eqnarray}
where
\begin{eqnarray}\label{Amp3D2}
&&{\tilde {\phi}}(\pm |p_x|,p_y,p_z)=\nonumber\\
&&\sqrt{\frac{|p_x|}{m}}~{\bar {\phi}}^\pm (\varepsilon=(p_x^2+p_y^2+p_z^2)/2m,p_y,p_z),
\end{eqnarray}
representing the probability amplitude for a free particle to arrive at $x=0$ with momentum ${\vec p}$.

Plugging Eqs.~(\ref{change}) and~(\ref{Amp3D}) into the solution of Eq.~(\ref{SolTYZ}), we obtain another representation of the three-dimensional free-particle solution in the STS extension:
\begin{equation}\label{Sol3D}
\begin{aligned}
&{\phi}^{\pm}(t,y,z|x)=\frac{1}{(2\pi\hbar)^{3/2}}\int_{-\infty}^{\infty}dp_ydp_z \int_{0}^{\infty} dp_x  \\
&\times \sqrt{\frac{p_x}{m}} ~{\tilde \phi}(\pm p_x,p_y,p_z)~ e^{i( p_y y+ p_z z-p^2 t/2m)/\hbar} e^{\pm i p_x x/\hbar},
\end{aligned}
\end{equation}
where $p^2={\vec p}\cdot{\vec p}$. Note that although $\phi(t,y,z|x)$ and $\psi(x,y,z|t)$ satisfy the same equation in the free-particle case, there is a clear distinction between Eq.~(\ref{Sol3D}) and the solution to the Schr\"odinger equation
\begin{eqnarray}\label{Sol3DQM}
\psi(x,y,z|t)
&=& \int dp_xdp_ydp_z ~{\tilde \psi}(p_x,p_y,p_z)\nonumber\\
&\times& e^{i( p_y y+ p_z z+ip_x x)/\hbar}~ e^{- i p^2t/2m\hbar }.
\end{eqnarray}
In addition, also note that in conventional QM, the counterpart to Eq.~(\ref{Amp3D}) is
\begin{eqnarray}\label{Amp3DQM}
{\tilde {\psi}}(p_x,p_y,p_z|t) = {\tilde {\psi}}(p_x,p_y,p_z) \, e^{-ip^2 t/2m\hbar},
\end{eqnarray}
which represents the probability amplitude for the particle to be found with momentum $\vec{p}$ at time $t$.

The probability density for the particle to arrive on the plane \( x = \text{constant} \) at time \( t \), regardless of its arrival position \( (y, z) \), is
\begin{eqnarray}\label{ProbX3D1}
&&{\cal P}_\phi(t|x)=\int_{-\infty}^{\infty} dydz~{\cal P}_\phi(t,y,z|x),
\end{eqnarray}
where  \( {\cal P}_\phi(t,y,z|x) \) is given by Eq.~(\ref{rhoTYZ}).
Substituting Eq.~(\ref{Sol3D}) into Eq.~(\ref{ProbX3D1}), we obtain
\begin{equation}\label{ProbX3D}
\begin{aligned}
&{\cal P}_{\phi}(t|x)=\frac{1}{\langle \phi(x)|\phi(x) \rangle}~\sum_{r=\pm}\int_{-\infty}^{\infty}dp_ydp_z 
 \bigg| \int_{0}^{\infty} dp_x \\
 & \sqrt{\frac{p_x}{2\pi m\hbar}} ~{\tilde \phi}(rp_x,p_y,p_z) ~ e^{-ip_x^2t/(2m\hbar)}~e^{irp_x x/\hbar} \bigg|^2. \\
\end{aligned}
\end{equation}
Similar to the one-dimensional case, Eq.~(\ref{ProbX3D}) can be identified with the traditional (normalized) Kijowski TOA distribution~\cite{Kijo} if one assumes \( {\tilde \phi}(p_x,p_y,p_z) = {\tilde \psi}(p_x,p_y,p_z) \). Remarkably, the Kijowski distribution—originally derived from a set of axioms intended to characterize a TOA distribution—emerges here from first principles, specifically from the SC Schr\"odinger equation~(\ref{Schro2TYZ}). It is also worth mentioning that Allcock derived Eq.~(\ref{ProbX3D}) using a phenomenological model based on a complex absorbing potential~\cite{All}.

Nevertheless, assuming \( {\tilde \phi}(p_x,p_y,p_z) = {\tilde \psi}(p_x,p_y,p_z) \) requires caution. While \( {\tilde \phi}(p_x,p_y,p_z) \) corresponds to the probability amplitude for the particle to arrive at \( x = 0 \) with momentum \( \vec{p} \), regardless of the arrival time, \( {\tilde \psi}(p_x,p_y,p_z) \) represents the amplitude for the particle to have momentum \( \vec{p} \) at time \( t = 0 \), regardless of its position. In fact, unlike \( {\tilde \phi}(p_x,p_y,p_z) \), \( {\tilde \psi}(p_x,p_y,p_z) \) is the spatial Fourier transform of \( \psi(x,y,z|t = 0) \), encoding delocalized spatial information about the momentum. Still, since \( {\tilde \psi}(p_x,p_y,p_z) \) is commonly used to predict arrival times in semiclassical regimes, we expect \( {\tilde \psi}(p_x,p_y,p_z) \approx {\tilde \phi}(p_x,p_y,p_z) \) in such contexts.

Even without direct access to the actual form of \( {\tilde \phi}(p_x,p_y,p_z) \), Eq.~(\ref{ProbX3D}) can still be tested experimentally. For example, by fixing the detector at \( x = 0 \), one can measure the particle’s arrival time distribution and infer the function \( {\bar \phi}(p_x,p_y,p_z) \) that best fits the data via Eq.~(\ref{ProbX3D}). If this same function continues to reproduce the experimental distribution \( {\cal P}_\phi(t|x) \) at different values of \( x \), this indicates that the STS extension provides a consistent prediction for the arrival time of a free particle.

\section{STS QM: unified four-dimensional formulation}\label{Unify}

In Cartesian coordinates, the arrival surface is defined by choosing one of the coordinates $x^i \in \{x, y,z\}$, which becomes the parameter of the quantum state $|\phi^i(x^i)\rangle$. For instance, when $x^i = x$, we have $|\phi^1(x^1)\rangle \equiv |\phi_x(x)\rangle$, with $\langle t,y,z|\phi_x(x)\rangle = \phi_x(t,y,z|x)$ representing the scenario analyzed in the previous sections. In the three-dimensional STS extension, the convenient choice of evolution parameter depends on the arrival surface under consideration---each choice giving rise to a different wave function of the same particle. 

Using Cartesian coordinates and incorporating conventional QM, the possible quantum states of the particle are
\begin{eqnarray}\label{states}
|\phi^0(x^0)\rangle &=& |\phi_t(t)\rangle \in {\cal H}|_T = {\cal H}_X \otimes {\cal H}_Y \otimes {\cal H}_Z,\nonumber\\
|\phi^1(x^1)\rangle &=& |\phi_x(x)\rangle \in {\cal H}|_X = {\cal H}_T \otimes {\cal H}_Y \otimes {\cal H}_Z,\nonumber\\
|\phi^2(x^2)\rangle &=& |\phi_y(y)\rangle \in {\cal H}|_Y = {\cal H}_T \otimes {\cal H}_X \otimes {\cal H}_Z,\nonumber\\
|\phi^3(x^3)\rangle &=& |\phi_z(z)\rangle \in {\cal H}|_Z = {\cal H}_T \otimes {\cal H}_X \otimes {\cal H}_Y.\nonumber\\
\end{eqnarray}
The dynamical equations for these states can be unified as
\begin{equation}\label{General}
{\hat P}^{\mu}\ket{{\phi}^{\mu}(x^{\mu})} = i\hbar~\eta^{\mu\nu}\frac{d}{dx^{\nu}}\ket{{\phi}^{\mu}(x^{\mu})},
\end{equation}
where $\eta^{\mu\nu} = \text{diag}(1,-1,-1,-1)$ and
\begin{eqnarray}\label{P}
{\hat P}^0 &=& {\hat H} = \frac{1}{2m} \left({\hat p}_x^2 + {\hat p}_y^2 + {\hat p}_z^2 \right) + V(t,{\hat x},{\hat y},{\hat z}),\nonumber\\
{\hat P}^1 &=& {\hat P}_x = \sigma_z \sqrt{2m \left[ {\hat h} - V({\hat t},x,{\hat y},{\hat z}) \right] - \left({\hat p}_y^2 + {\hat p}_z^2\right)},\nonumber\\
{\hat P}^2 &=& {\hat P}_y = \sigma_z \sqrt{2m \left[ {\hat h} - V({\hat t},{\hat x},y,{\hat z}) \right] - \left({\hat p}_x^2 + {\hat p}_z^2\right)},\nonumber\\
{\hat P}^3 &=& {\hat P}_z = \sigma_z \sqrt{2m \left[ {\hat h} - V({\hat t},{\hat x},{\hat y},z) \right] - \left({\hat p}_x^2 + {\hat p}_y^2\right)}.\nonumber\\
\end{eqnarray}

We can go beyond Eq.~(\ref{General}) and unify the STS extension and conventional QM within a timeless and spaceless framework. As discussed in the introduction, a fully quantum description of time that recovers the Schr\"odinger equation (where time is treated as a parameter) is achieved via Wheeler--DeWitt-type equations~\cite{Rovelli,Smith,Page,Gio,Vedral}—timeless constraints satisfied by static quantum states. In this formalism, $|\psi(t)\rangle$ emerges by conditioning (i.e., projecting) a spacetime-independent state onto an eigenstate $|t\rangle$.

Accordingly, the state $|\phi_x(x)\rangle$, for example, should emerge analogously to $|\psi(t)\rangle$, by conditioning a space-independent state onto an eigenstate $|x\rangle$. In this case, since position plays the role of the evolution parameter, we must seek a fully quantum description of space via a spaceless constraint. To this end, we will incorporate into the framework the eigenstate $|x\rangle$ from conventional QM, which does not appear in the STS extension when $x$ is chosen as the evolution parameter.

A Wheeler–DeWitt-type equation for conventional QM can be obtained by extending the phase space of the particle, where $(p_0, p_1, p_2, p_3) := (p_t = E, p_x, p_y, p_z)$ are the conjugate momenta associated with the space-time coordinates $(x^0, x^1, x^2, x^3) = (t, x, y, z)$, defined with respect to an inertial observer~\cite{Rovelli,Smith}. Consequently, the Hilbert space of the particle is extended to ${\cal H}_{\rm total} := {\cal H}_T \otimes {\cal H}_X \otimes {\cal H}_Y \otimes {\cal H}_Z$. The quantization rule introduced in Eq.~(\ref{TYZ}), together with an analogous quantization for the $x$ coordinate, promotes the phase-space observables to operators ${\hat x}^\mu$ and ${\hat p}^\mu$ acting on ${\cal H}_{X^\mu}$ (with $X^\mu = T, X, Y$, or $Z$), satisfying the canonical commutation relations
\begin{eqnarray}\label{XTYZ}
[{\hat x}^\mu, {\hat p}^\nu] = -i\hbar~ \eta^{\mu\nu},
\end{eqnarray}
where ${\hat p}^0 = {\hat p}_t := {\hat h}$. We emphasize the distinction between the operator ${\hat P}^\mu$, defined in Eq.~(\ref{P}) as governing the evolution with respect to $x^\mu$, and ${\hat p}^\mu$, defined via its canonical relation with ${\hat x}^\mu$ in Eq.~(\ref{XTYZ}).

A timeless constraint that leads to the Schrödinger equation is obtained by quantizing the classical Hamiltonian constraint
\begin{eqnarray} \label{PWt}
{\mathbbm{P}}^0 := -{p}_t + \frac{1}{2m}\big({p}_x^2 + {p}_y^2 + {p}_z^2 \big) + V({t}, {x}, {y}, {z}) = 0.
\end{eqnarray}
Applying the quantization rule~(\ref{XTYZ}) to Eq.~(\ref{PWt}), we obtain the condition that a time-independent physical state $|\Phi^0\rangle$ must satisfy
\begin{eqnarray} \label{PWT}
\hat{\mathbbm{P}}^0 |\Phi^0\rangle := \big(-{\hat p}_t + {\hat{\hat P}}_t \big) |\Phi^0\rangle = 0,
\end{eqnarray}
where
\begin{eqnarray} \label{HT}
{\hat {\hat  P}}_t := \frac{1}{2m}\big({\hat p}_x^2 + {\hat p}_y^2 + {\hat p}_z^2 \big) + V({\hat t}, {\hat x}, {\hat y}, {\hat z}).
\end{eqnarray}
The distinction between ${\hat{\hat P}}_t$ and ${\hat P}_t = {\hat H}$ [as defined in Eq.~(\ref{P})] lies in the treatment of time: in ${\hat{\hat P}}_t$, time is a quantum operator, whereas in ${\hat P}_t$ it is a parameter.

Any solution of Eq.~(\ref{PWT}) can be expressed as a complete history of the particle in the form
\begin{eqnarray} \label{PWTsol}
\ket{\Phi^0} = \int dt ~ |t\rangle \otimes |\phi_t(t)\rangle,
\end{eqnarray}
where $|\phi_t(t)\rangle$ represents the state of the particle at time $t$. Explicitly,
\begin{align}\label{stateT}
\ket{\phi_t(t)} =   \langle t|\Phi^0\rangle ~~\in  ~~{\cal H}_{X} \otimes {\cal H}_{Y} \otimes {\cal H}_{Z}.
\end{align}
Note that $|\Phi^0\rangle$ is not a proper element of ${\cal H}_{\rm total}$, being normalized with respect to the inner product
\begin{equation} \label{innerproductT}
( \Phi^0|\Phi^0)_T = \langle \Phi^0| \Big(|t\rangle \langle t| \otimes \mathbbm{1}_{XYZ} \Big) |\Phi^0 \rangle = \langle \phi_t(t)|\phi_t(t)\rangle = 1,
\end{equation}
for all $t \in \mathbbm{R}$. The formalism presented in Eqs.~(\ref{PWT})–(\ref{innerproductT}) constitutes a well-established timeless formulation of convencional QM~\cite{Rovelli,Smith}.

To verify that $|\phi_t(t)\rangle$ satisfies the Schrödinger equation, we insert Eq.~(\ref{PWTsol}) into Eq.~(\ref{PWT}) and project the resulting expression onto $|t\rangle$, yielding
\begin{eqnarray} \label{PWTproj}
\int dt' ~\Big[-\langle t|{\hat p}_t|t'\rangle  |\phi_t(t')\rangle + \langle t| {\hat {\hat P}}_t \Big(|t'\rangle \otimes |\phi_t(t')\rangle \Big) \Big] = 0. \nonumber\\
\end{eqnarray}
Using $\langle t|{\hat p}_t|t'\rangle = i\hbar \frac{d}{dt}\delta(t-t')$ and
\begin{eqnarray} \label{H}
\int dt' ~\langle t|{\hat {\hat P}}_t\Big(|t'\rangle \otimes |\phi_t(t')\rangle \Big) = {\hat P}_t|\phi_t(t)\rangle,
\end{eqnarray}
where ${\hat P}_t = {\hat H}$, Eq.~(\ref{PWTproj}) leads directly to the Schrödinger equation:
\begin{equation} \label{SE}
{\hat P}_t|\phi_t(t)\rangle = i\hbar \frac{d}{dt}|\phi_t(t)\rangle.
\end{equation}

Note that as ${\hat p}_t$ is the generator of time translations in ${\cal H}_T$, the constraint in Eq.~(\ref{PWT}) implies that ${\hat  P}_t$ governs the evolution of the degrees of freedom in ${\cal H}_X \otimes {\cal H}_Y \otimes {\cal H}_Z$ with respect to the time coordinate $x^0 = t$. A similar construction can be applied to derive a spaceless framework leading to the STS extension. Choosing $x$ as the parameter, the Wheeler–DeWitt equation must now contain ${\hat p}_x$, which is the generator of space translations in ${\cal H}_X$.

To do this, we start—analogously to Eq.~(\ref{PWT})—from the classical constraint in Eq.~(\ref{PWt}), but now isolating the momentum in the $x$ direction:
\begin{eqnarray} \label{PWx}
P^1 := -p_x \pm \sqrt{2m \big[p_t - V(t,x,y,z)\big] - (p_y^2 + p_z^2)} = 0.\nonumber\\
\end{eqnarray}
Although Eqs.~(\ref{PWt}) and~(\ref{PWx}) are classically equivalent, their quantizations lead to different quantum theories. The quantization of Eq.~(\ref{PWx}) gives
\begin{eqnarray} \label{PWX}
\hat{\mathbbm{P}}^1 |\Phi^1\rangle := \big( -{\hat p}_x + {\hat {\hat P}}_x \big) |\Phi^1\rangle = 0,
\end{eqnarray}
where
\begin{eqnarray} \label{PC}
{\hat {\hat P}}_x := \sigma_z \sqrt{2m \big[{\hat p}_t - V({\hat t},{\hat x},{\hat y},{\hat z})\big] - ({\hat p}_y^2 + {\hat p}_z^2)}.
\end{eqnarray}
Since $[{\hat p}_x, {\hat {\hat P}}_x] \neq 0$, we cannot recover the traditional constraint of Eq.~(\ref{PWT}). As anticipated in the previous sections, Eqs.~(\ref{PWT}) and~(\ref{PWX}) correspond to distinct quantum constraints for the same particle.

The solution of Eq.~(\ref{PWX}) takes the form
\begin{eqnarray} \label{PWXSol}
|\Phi^1\rangle = \int dx ~ |\phi_x(x)\rangle \otimes |x\rangle,
\end{eqnarray}
which serves as the spatial counterpart of the well-known timeless state given in Eq.~(\ref{PWTsol}). Here,
\begin{align}\label{stateX}
|\phi_x(x)\rangle = \langle x|\Phi^1\rangle ~~\in~~ {\cal H}_T \otimes {\cal H}_Y \otimes {\cal H}_Z
\end{align}
represents the quantum state of the particle conditioned on position $x$.  As in the temporal case, $|\Phi^1\rangle$ is not a proper element of ${\cal H}_{\rm total}$, but it is normalized with respect to the inner product
\begin{eqnarray} \label{innerproductT}
(\Phi^1|\Phi^1)_X &=& \langle \Phi^1| \big( |x\rangle \langle x| \otimes \mathbbm{1}_{TYZ} \big) |\Phi^1 \rangle \nonumber\\
&=& \langle \phi_x(x)|\phi_x(x)\rangle = 1.
\end{eqnarray}

Substituting Eq.~(\ref{PWXSol}) into Eq.~(\ref{PWX}) and projecting onto $|x\rangle$, we obtain
\begin{eqnarray} \label{PWTsub}
\int dx' ~\Big[-\langle x|{\hat p}_x|x'\rangle |\phi_x(x')\rangle + \langle x| {\hat {\hat P}}_x \big(|x'\rangle \otimes |\phi_x(x')\rangle \big) \Big] = 0. \nonumber\\
\end{eqnarray}
Using $\langle x|{\hat p}_x|x'\rangle = -i\hbar \frac{d}{dx}\delta(x - x')$ and
\begin{eqnarray} \label{Px}
\int dx'~ \langle x|{\hat {\hat P}}_x\big(|x'\rangle \otimes |\phi_x(x')\rangle \big) = {\hat P}_x |\phi_x(x)\rangle,
\end{eqnarray}
with ${\hat P}_x$ defined in Eq.~(\ref{P}), Eq.~(\ref{PWTsub}) yields the space-conditional (SC) Schrödinger equation~(\ref{SchroT}) for the three-dimensional case.

Given the above, we readily see that the constraints associated with each state $|\Phi^\mu\rangle$ (with $\mu = 0, 1, 2, 3$) is
\begin{eqnarray} \label{PWG}
\hat{\mathbbm{P}}^\mu |\Phi^\mu\rangle = 0, \quad \text{with} \quad \hat{\mathbbm{P}}^\mu := -{\hat p}^\mu + {\hat{\hat P}^\mu}.
\end{eqnarray}
The corresponding solutions admit the expression
\begin{eqnarray} \label{PWGsol}
|\Phi^\mu\rangle = \int dx^\mu ~ |\phi^\mu(x^\mu)\rangle \otimes |x^\mu\rangle,
\end{eqnarray}
where $|\Phi^\mu\rangle \in {\cal H}_{\rm total} = \bigotimes_{\mu = 0}^3 {\cal H}_{X^\mu}$, $|x^\mu\rangle \in {\cal H}_{X^\mu}$, and $|\phi^\mu(x^\mu)\rangle \in \bigotimes_{\mu' \neq \mu} {\cal H}_{X^{\mu'}}$. 

Projecting Eq.~(\ref{PWG}) onto $|x^\mu\rangle$, and using the relations
\begin{equation} \label{p}
\langle x^\mu |{\hat p}^\mu|\Phi^\mu\rangle = i \hbar~ \eta^{\mu\nu} \frac{\partial}{\partial x^\nu} \langle x^\mu|\Phi^\mu\rangle,
\end{equation}
and
\begin{equation} \label{Pmu}
\langle x^\mu |{\hat{\hat P}^\mu}|\Phi^\mu\rangle = {\hat P}^\mu \langle x^\mu|\Phi^\mu\rangle,
\end{equation}
we recover the generalized Schrödinger-type equation presented in Eq.~(\ref{General}). Each $|\Phi^\mu\rangle$ satisfies a normalization condition specific to its parameter:
\begin{eqnarray} \label{innerproductX}
(\Phi^\mu|\Phi^\mu)_{X^\mu} &=& \langle \Phi^\mu| \Big(|x^\mu\rangle \langle x^\mu| \otimes \mathbbm{1} \Big) |\Phi^\mu \rangle \nonumber\\
&=& \langle \phi^\mu(x^\mu)|\phi^\mu(x^\mu)\rangle = 1.
\end{eqnarray}

Finally, note that $|\Phi^0\rangle$ describes the joint probability amplitudes of the spatial coordinates $\{x^i\}$ along the time-like direction $x^0 = t$, capturing the particle’s history. In contrast, each $|\Phi^i\rangle$ encodes the joint probability amplitudes of $\{x^\mu\}_{\mu \neq i}$ along the space-like direction $x^i$. This formalism naturally suggests a relativistic generalization. To that end, it is important to note that separating the time coordinate from the spatial ones via a tensor product structure implicitly presupposes a spacetime foliation.

\section{Conclusion}
\label{conclusion}

We have generalized the STS extension of quantum mechanics to three spatial dimensions, introducing space-conditional wave functions such as \( \phi(t, y, z | x) \), which complement the conventional time-conditional (Schr\"odinger) wave function \( \psi(x, y, z | t) \). This extension restores the symmetry between space and time in the non-relativistic quantum description, allowing different coordinates to serve as ``evolution'' parameters.

These wave functions are encoded in generalized states \( |\phi^\mu(x^\mu)\rangle \), where each choice of \( x^\mu \in \{t, x, y, z\} \) defines a distinct description of the system on a given hypersurface. For \( x^\mu = t \), the STS framework recovers standard non-relativistic quantum mechanics. For \( x^\mu = x \), the state \( |\phi_x(x)\rangle \) leads to the conditional wave function \( \phi(t, y, z | x) \), which predicts the position and time of arrival on the plane \( x = \text{const} \). In the free-particle case, we have verified that \( \phi(t, y, z |x) \) naturally reproduces the mathematical form of the well-known axiomatic Kijowski distribution.

Importantly, we have shown that the different conditional states \( |\phi^\mu(x^\mu)\rangle \) can emerge as projections of a timeless and spaceless physical state, satisfying different constraints of the form \( \hat{\mathbbm{P}}^\mu |\Phi^\mu\rangle = 0 \). This formulation generalizes the spirit of the Wheeler–DeWitt equation: rather than selecting time as a privileged evolution parameter, all coordinates are treated on equal footing. It suggests that the STS extension encodes complementary information about where and when arrival events are expected to occur, even in scenarios beyond standard quantum mechanics.

Future investigations may explore how ideal limits of realistic measurements involving clocks and detectors, described within conventional QM, can be recovered in a manner consistent with the STS framework, and how the STS formalism might be embedded within a broader relativistic or quantum field-theoretic context.

\section{\label{sec:level8}acknowledgements}
I am grateful to Fernando Parisio, Ricardo Ximenes, and Pedro Medeiros for the fruitful discussions that helped shape the ideas presented in this work.

\end{document}